\documentclass[aps,prl,twocolumn,floatfix,longbibliography,nofootinbib]{revtex4-2}

\usepackage[utf8]{inputenc}
\usepackage{natbib}
\usepackage{graphicx}
\usepackage[percent]{overpic}
\usepackage{xcolor}
\usepackage[tbtags]{amsmath}
\usepackage{amssymb}
\usepackage{braket}
\usepackage{bm}
\usepackage{comment}
\usepackage{multirow}
\usepackage[normalem]{ulem}
\usepackage{xr-hyper}
\usepackage[colorlinks,allcolors=black,plainpages=false]{hyperref}

\makeatletter
\newcommand{\importmanuscriptlabels}[1]{%
  \externaldocument[][nocite]{#1}%
  \let\r@FirstPage\@undefined
  \let\r@LastPage\@undefined
  \let\r@LastBibItem\@undefined
}
\makeatother

\graphicspath{{../references/}{../figures/}{figures/}}

\renewcommand{\vec}[1]{\boldsymbol{#1}}
\def\vA{\vec{A}}
\def\vE{\vec{E}}
\def\vk{\vec{k}}
\def\vq{\vec{q}}
\def\vp{\vec{p}}

\def\vR{\vec{R}}
\def\vr{\vec{r}}
\def\ve{\varepsilon}
\def\dg{\dagger}
\def\nn{\nonumber}
\def\beq{\begin{eqnarray}}
\def\eeq{\end{eqnarray}}
\def\la{\langle}
\def\ra{\rangle}
\def\PP{{\cal P}}
\def\M{{\cal M}}
\def\Z{{\cal Z}}
\def\F{{\cal F}}
\def\E{{\cal E}}
\def\N{{\cal N}}

\newif\ifincludesupplement
\ifdefined\OMITSUPPLEMENT
  \includesupplementfalse
\else
  \includesupplementtrue
\fi
\ifincludesupplement\else
  \importmanuscriptlabels{supp}
\fi

\begin{document}

\title{Stability, Collective Modes, and Quasiparticle--Phonon Coupling of Metallic Wigner Crystals in Rhombohedral Graphene}

\author{Sunghoon Kim}
\author{Yang-Zhi Chou}
\author{Jay D. Sau}
\author{Sankar Das Sarma}
\affiliation{Condensed Matter Theory Center and Joint Quantum Institute,
Department of Physics, University of Maryland, College Park, MD 20742, USA}

\begin{abstract}
Recent transport experiments in rhombohedral multilayer graphene have reported evidence for metallic Wigner crystals (mWCs), characterized by the coexistence of crystalline order and a small density of hole-like carriers. This coexistence raises fundamental questions about the stability of mWCs and the fate of their collective excitations in the presence of gapless fermions. Using self-consistent Hartree–Fock and time-dependent Hartree–Fock calculations, we demonstrate that itinerant carriers profoundly influence both. We find that mWCs with electron pockets are generally unstable to particle–hole fluctuations, while hole-doped mWCs at lower densities remain locally stable. In the stable hole-doped mWC, the longitudinal phonon strongly hybridizes with particle–hole excitations, while the transverse phonon remains relatively well defined. Quasiparticle coupling is strongly mode-dependent and is typically stronger for longitudinal-like excitations than for transverse modes. Our results demonstrate that self-doped carriers not only enable metallicity but also fundamentally reshape the stability, collective dynamics, and quasiparticle interactions of electronic crystals.
\end{abstract}

\maketitle

\textit{Introduction.}---
Wigner crystallization is a paradigmatic example of interaction-driven spontaneous symmetry breaking \cite{Wigner_PR1934}. While conventional WCs have one electron per crystalline unit cell, electronic crystals can also form at incommensurate fillings and remain metallic. Such a \textit{metallic Wigner crystal} (mWC) has been proposed as a possible intermediate phase between a commensurate WC and a uniform Fermi liquid, arising from the self-doping instability of the WC \cite{Fisher-Halperin-Morf_PRB79,Cockayne-Elser_PRB91,Candido_PRL01,Falakshahi_PRL05,Kim-Kivelson_PRB24}. The coexistence of crystalline order and gapless fermions raises fundamental questions about the stability of the crystal against particle--hole fluctuations and the fate of the Goldstone modes \cite{Tesanovic-Halperin_PRB89,Cote-MacDonald_PRB91} when coupled to itinerant fermions \cite{Kivelson-Fradkin_PRB01, Sun-Fradkin_PRB08}.

Rhombohedral multilayer graphene (RMG) provides a promising platform for studying electronic crystallization beyond conventional two-dimensional electron gases \cite{Geier_NatCommun26}. Its tunable bands exhibit nontrivial quantum geometry and develop a Mexican-hat dispersion under a strong displacement field \cite{Zhang-Min-MacDonald_PRB10}. The interplay of this tunable electronic structure and Coulomb interactions is thought to play an important role in the emergence of various correlated phases, including spin and valley polarized metals \cite{Zhou_Nature21a}, integer and fractional quantum anomalous Hall states \cite{Lu_Nature24,Lu_Nature25,Choi_Nature25}, and superconductivity \cite{Zhou_Nature21b,Han_Nature25,Kumar_Nat2026-DualSC,Qin2026-stripeSC,Deng2026-cornellSC,Kalantre2026-qoSC,Hua2026-multiknobSC,Zheng2026-R5GCSC}. More recently, transport experiments have reported evidence for WCs and mWCs at low electron densities \cite{Han-Ju_26,zhou2026expt}, near the putative chiral superconducting region \cite{Han_Nature25}. In particular, a narrow metallic regime adjacent to the WC phase exhibits a small density of hole-like carriers, consistent with a hole-doped mWC. Independently, scanning tunneling microscopy has reported evidence for mWCs in a distinct parameter regime \cite{Sun-Liu_26}.

Motivated by these experiments, recent theories have studied the physics of WCs and mWCs in RMG \cite{Dong-Vishwanath_arxiv26,Feng-Zaletel_arxiv26,Abouelkomsan-Fu_arxiv26,Kim-Wen_arxiv26,Ge-Yu_arxiv26,guerci_arxiv2026,Miao-Huang_arxiv2026,Babbar_PRB26}. In particular, self-consistent Hartree--Fock (HF) calculations predict substantial mWC regions with hole pockets, electron pockets, or both, whereas transport evidence has so far been concentrated in a relatively narrow hole-doped mWC regime \cite{Han-Ju_26}. This raises the question of whether the predicted mWC states remain stable against fluctuations of their Fermi pockets \cite{Jain-Huang_PRB26}. Even when the mWC state is locally stable, its gapless fermions may strongly modify collective modes. Although symmetry constrains the coupling between the quasiparticle and the translational Goldstone mode in symmetry-broken metals to vanish in the long-wavelength limit \cite{Watanabe-Vishwanath_PNAS2014}, whether well-defined phonons survive at finite momenta, and how they interact with itinerant fermions, require microscopic investigation.

\begin{figure*}[pth!]
\centering
\includegraphics[width=0.8\linewidth]{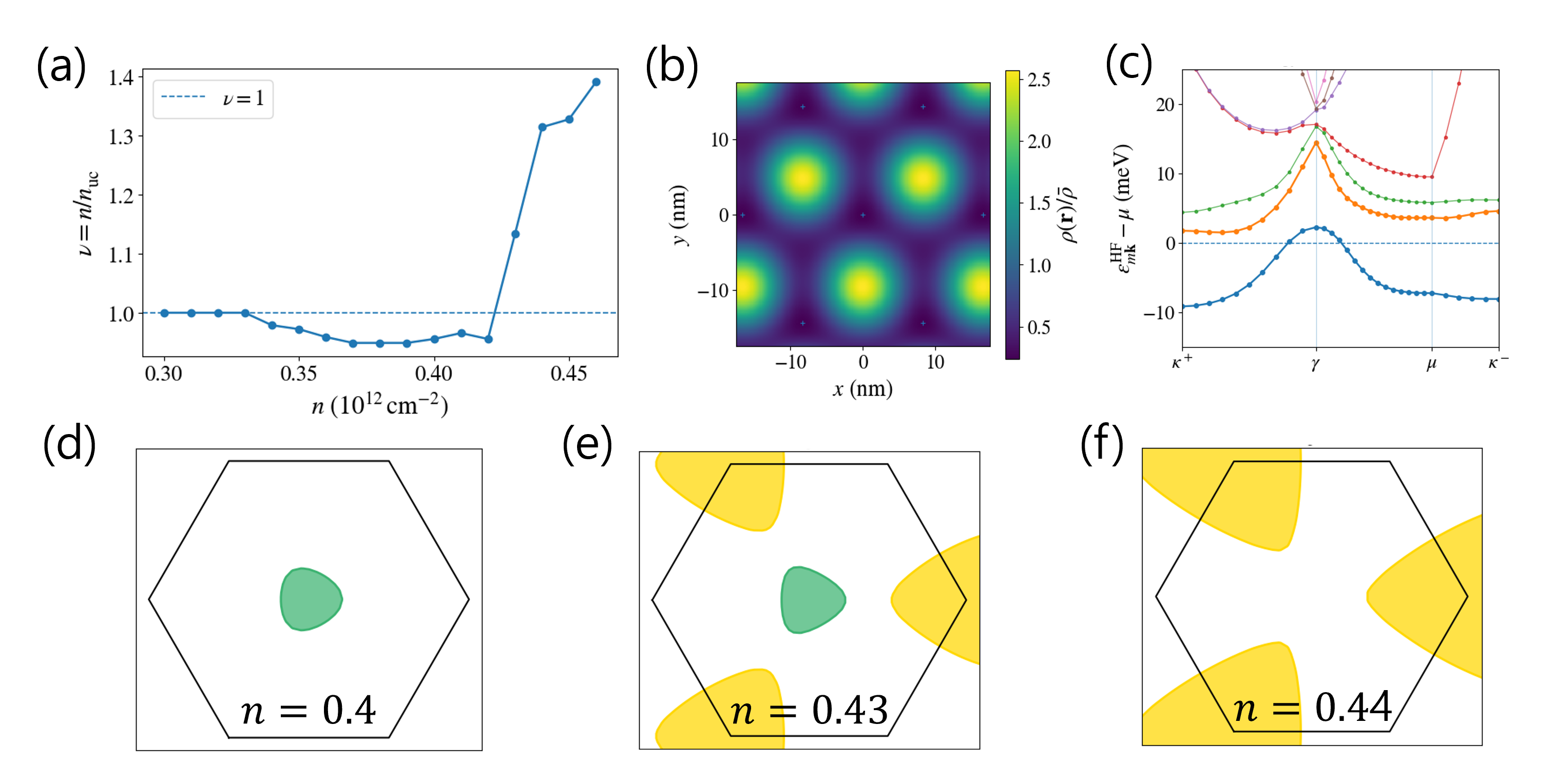} 
\caption{\textbf{Self-consistent HF solutions and Fermi-pocket evolution of mWCs.}
(a) Optimal filling $\nu=n/n_{\rm uc}$ as a function of electron density $n$, obtained by minimizing the HF energy per electron at each fixed $n$. The dashed line indicates the commensurate filling $\nu=1$ (i.e., WC). Upon increasing $n$, $\nu$ deviates from unity, indicating the emergence of incommensurate mWCs. (b) Real-space charge density and (c) low-energy HF band structure for the hole-doped mWC at $n=0.4\times10^{12}\mathrm{cm}^{-2}$.
(d)--(f) Fermi pockets at three representative densities, $n=(0.4,0.43,0.44)\times10^{12}\mathrm{cm}^{-2}$. Green and yellow shaded regions denote hole and electron pockets, respectively, and black hexagons indicate the crystalline Brillouin zones (BZs). The stability analysis (Fig.~\ref{fig:Stability}) shows that mWC solutions with electron pockets (including (e) and (f)) are unstable. 
}
\label{fig:HF}
\end{figure*}

In this Letter, we investigate the stability and collective modes of mWCs in RMG, as well as their coupling to quasiparticles. We begin with self-consistent HF results showing mWCs over a broad range of electron density. By analyzing the stability matrix, we find that the mWCs with electron pockets generally become unstable through intra- and interband particle--hole fluctuations, while hole-doped mWCs at lower densities remain locally stable, consistent with the recent transport experiments \cite{Han-Ju_26,zhou2026expt}. In the stable hole-doped regime, we compute collective excitations using time-dependent Hartree--Fock (TDHF) calculations. Using various quantitative diagnostics, we find that the longitudinal mode strongly mixes with the particle--hole continuum, while the transverse phonon mode remains relatively clean. Finally, at a representative quasiparticle momentum near the Fermi level, we calculate the coupling to TDHF excitations. The coupling shows a strong mode dependence and is typically stronger for mixed longitudinal-like modes than for transverse-like modes, highlighting the importance of the full spectrum of particle--hole excitations. Our work demonstrates how metallicity fundamentally alters the stability and collective dynamics of electronic crystals.

\textit{Model.}---
We consider spin- and valley-polarized electrons in RMG with screened Coulomb interactions,
\beq
H=H_0
+\frac{1}{2A}\sum_{\vq}V(q):\rho_{\vq}\rho_{-\vq}:,
\eeq
where $H_0$ describes a tight-binding RMG Hamiltonian with a perpendicular displacement field and $A$ is the system area. We use the screened Coulomb interaction $V(q)=\frac{e^2}{2\epsilon_0\epsilon_r q }\tanh(qd)$, where $\epsilon_0$ is vacuum permittivity, $\epsilon_r$ is the relative dielectric constant, and $d$ is the distance to the metallic gate. The density operator in the noninteracting RMG band basis, with band indices $a,b$, is $\rho_{\vq}=\sum_{\vp,a,b}\langle u_a(\vp+\vq)|u_b(\vp)\rangle c^\dagger_{\vp+\vq,a}c_{\vp,b}$.

To study electron crystals in the interacting RMG model, we perform HF calculations, focusing on triangular lattices, which is a common crystalline structure of WCs, with electron density $n$ that may be incommensurate with the unit-cell density $n_{\rm uc}$. The HF state is characterized by the one-body density matrix $P_{\alpha\beta}(\vk)=\langle d^\dagger_{\vk\alpha}d_{\vk\beta}\rangle$, where $\alpha,\beta$ label the zone-folded states of the projected parent band, and $\vk$ denotes the reduced momentum in the crystalline Brillouin zone (BZ) measured relative to $K$. The self-consistent HF Hamiltonian is $h[P]=h_0+\Sigma_{\rm H}[P]+\Sigma_{\rm F}[P]$, where $\Sigma_{\rm H}$ and $\Sigma_{\rm F}$ denote the Hartree and Fock terms, respectively. For the self-consistent solution $P_0$, we obtain the reconstructed HF bands $\epsilon_m(\vk)$ and quasiparticle operators $\gamma_{\vk m}=\sum_\alpha C_{\alpha m}^*(\vk)d_{\vk\alpha}$. We determine the optimal $n_{\rm uc}$ at each fixed $n$ by minimizing the HF energy per electron. In numerical calculations, we focus on tetralayer RMG with an interlayer potential difference $U=60$ meV, $\epsilon_r=15$, $d=25$ nm, and $N_c = 30\times30$ crystalline unit cells. Details of the microscopic Hamiltonian and HF calculations are provided in the Supplemental Material \cite{si}.

\begin{figure*}[pth!]
\centering
\includegraphics[width=0.8\linewidth]{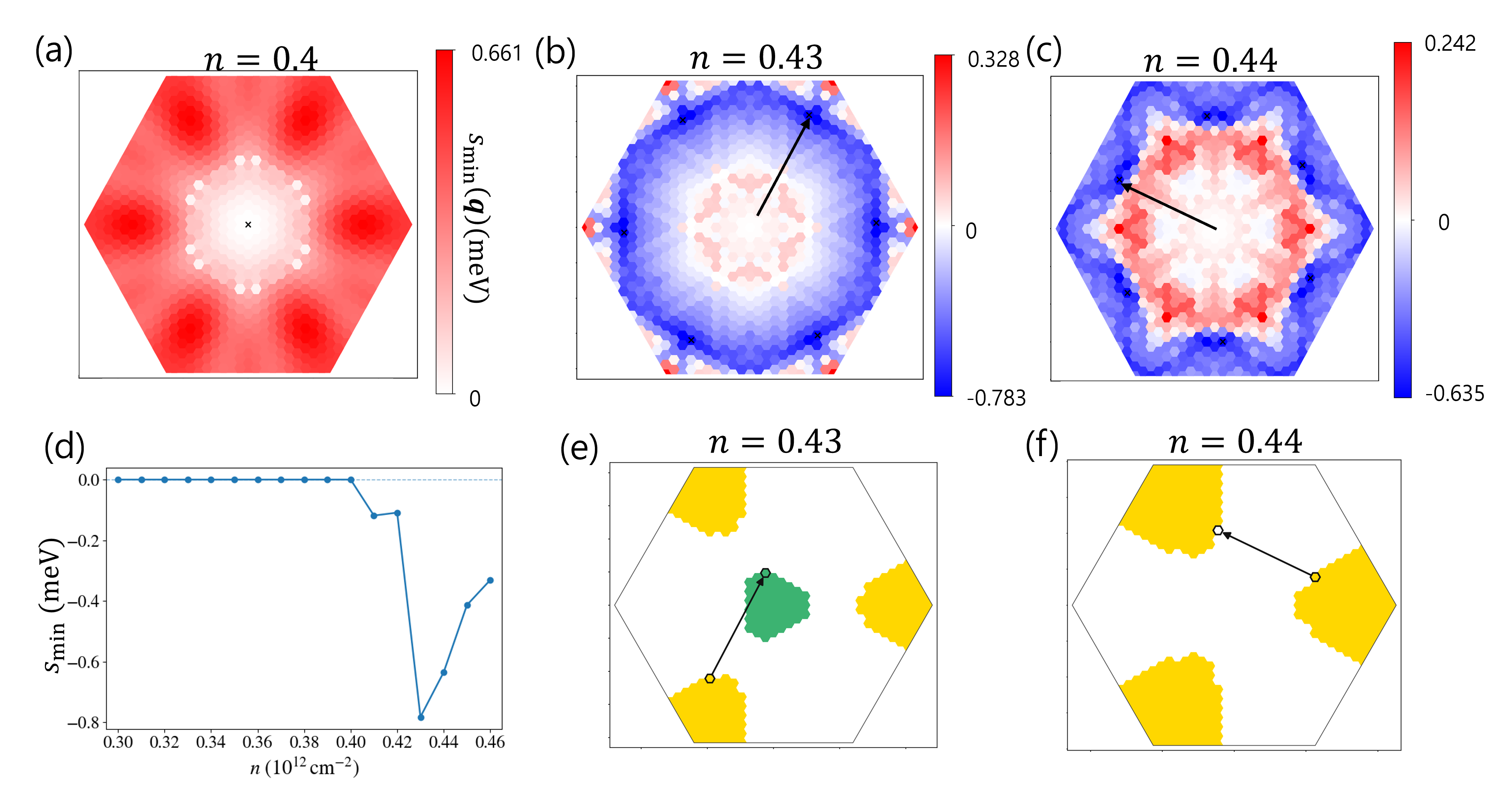} 
\caption{\textbf{Stability analysis of mWCs.}
(a--c) Lowest eigenvalue $s_{\min}(\vq)$ of the stability matrix $S(\vq)$ across the BZ at $n=(0.4,0.43,0.44)\times10^{12}\,\mathrm{cm}^{-2}$, respectively. The hole-doped mWC in (a) is locally stable, whereas mWCs with electron pockets in (b,c) exhibit instabilities at finite $\vq$. Black arrows indicate most unstable momenta. (d) Minimum eigenvalue $s_{\min}=\min_{\vq}s_{\min}(\vq)$ as a function of $n$. The dashed line indicates zero eigenvalue.
(e,f) Dominant particle--hole fluctuations associated with the most unstable eigenvectors in (b,c), respectively. Green and yellow shaded regions denote hole and electron pockets, respectively, with occupations of 0 and 1.
\label{fig:Stability}}
\end{figure*}

\textit{HF results.}---
Fig.~\ref{fig:HF}(a) summarizes the evolution of the optimal filling $\nu=n/n_{\rm uc}$ as a function of $n$. At low densities $n\lesssim0.33\times10^{12}\mathrm{cm}^{-2}$, the commensurate WC with $\nu=1$ is energetically favored. Upon increasing $n$, $\nu$ deviates from unity, indicating self-doping of the WC and the emergence of incommensurate mWCs. At lower densities, $\nu$ is slightly below unity, indicating a hole-doped mWC. At higher densities, $\nu$ increases above unity, indicating an electron-doped mWC.

The mWCs exhibit strong modulation of the real-space charge density, as shown in Fig.~\ref{fig:HF}(b). Unlike the commensurate WC, however, the mWC hosts partially filled HF bands [Fig.~\ref{fig:HF}(c)], indicating the coexistence of itinerant carriers with the underlying crystal. Figs.~\ref{fig:HF}(d)--(f) illustrate the evolution of the Fermi pockets with increasing $n$. At lower densities $n\lesssim0.42\times10^{12}\mathrm{cm}^{-2}$, the mWC hosts a hole pocket centered at $\gamma$. At higher densities $n\gtrsim0.44\times10^{12}\mathrm{cm}^{-2}$, the mWC features electron pockets at the zone corners of the crystalline BZ. There is a narrow intermediate regime where hole and electron pockets coexist. Importantly, the net self-doping in the intermediate regime reflects the difference between the electron- and hole-pocket areas, rather than their sum. These results are qualitatively consistent with the previous theory work \cite{Dong-Vishwanath_arxiv26}.

\textit{Stability of mWCs.}---
While the optimal $n_{\rm uc}$ minimizes the HF energy per electron within the triangular lattice ansatz, the resulting solution may be a saddle point of the full HF energy functional rather than a local minimum. We therefore examine the stability of the mWC solutions by analyzing the curvature of the HF energy functional \cite{Thouless_1960}. We perturb the self-consistent HF solutions by mixing occupied and unoccupied HF states with momentum transfer $\vq$ and expand the HF energy change to quadratic order:
\beq
\delta E_{\rm HF}^{(2)}
=\frac{1}{2N_c}\sum_{\vq}
\vec{v}_{\vq}^\dg S(\vq)\vec{v}_{\vq},
\eeq
where $S(\vq)$ is the stability matrix. Each eigenvector $\vec{v}_{\vq}$ represents a superposition of transitions from occupied to unoccupied HF states carrying momentum $\vq$. A negative eigenvalue of $S(\vq)$ identifies a fluctuation lowering the HF energy, indicating the instability of the self-consistent solution \cite{si}.

Figs.~\ref{fig:Stability}(a)--(c) show the lowest eigenvalue $s_{\min}(\vq)$ across the BZ for three representative mWCs corresponding to those shown in Figs.~\ref{fig:HF}(d)--(f). The hole-doped mWC has nonnegative eigenvalues throughout the BZ [Fig.~\ref{fig:Stability}(a)]. The minimum eigenvalue vanishes at $\gamma$, reflecting the gapless Goldstone modes. This indicates that the hole-doped mWC is locally stable. In contrast, the mWCs with electron pockets show negative eigenvalues at finite $\vq$ [Figs.~\ref{fig:Stability}(b), (c)], indicating instabilities against finite-$\vq$ particle--hole fluctuations. In Fig.~\ref{fig:Stability}(d), we plot the minimum eigenvalue $s_{\min}\equiv\min_{\vq}s_{\min}(\vq)$ as a function of $n$. The commensurate WCs and low-density hole-doped mWCs are stable, while higher-density mWCs with electron pockets are unstable, developing negative $s_{\rm min}$. In the present calculation with $N_c=900$, the onset of instability appears on the high-density side of the hole-doped mWC regime, although its precise location exhibits system-size dependence \cite{si}. Remarkably, the stability of low-density hole-doped mWCs and the instability of electron-doped mWCs are qualitatively consistent with the recent transport experiments, which report only hole-like itinerant carriers in the narrow mWC region \cite{Han-Ju_26}. 

To identify the most unstable fluctuations, we focus on the eigenvectors with the most negative eigenvalues, and find transitions with the largest weight. Figs.~\ref{fig:Stability}(e), (f) illustrate the most unstable particle--hole fluctuations of the two representative unstable mWCs. At $n=0.43\times10^{12}\mathrm{cm}^{-2}$, the dominant instability arises from an interband particle--hole transition between an electron pocket and the hole pocket [Fig.~\ref{fig:Stability}(e)]. At $n=0.44\times10^{12}\mathrm{cm}^{-2}$, the dominant instability is associated with an intraband transition between distinct electron pockets [Fig.~\ref{fig:Stability}(f)]. These results demonstrate that the stability of the mWC depends on the Fermi-pocket distribution, and reveal how the itinerant carriers can destabilize the mWC state.

\textit{Collective excitations.}---
Next, we compute the collective excitations of the stable hole-doped mWC using TDHF. Linearizing the HF equations of motion gives $L_{\vq}\vec x_{\lambda\vq}=E_\lambda(\vq)\vec x_{\lambda\vq}$, where $L_{\vq}$ is the TDHF operator, $\lambda$ denotes the mode label, and $E_\lambda(\vq)$ is the spectrum. The eigenvector $\vec x_{\lambda\vq}$ consists of particle--hole amplitudes $x_{\lambda,\vk mn}(\vq) =\delta\braket{\gamma^\dg_{\vk m}\gamma_{\vk+\vq,n}}_\lambda$ in the HF basis.

In mWCs, the crystalline order and itinerant carriers arise from the same interacting electronic system, rather than two independent sectors supporting separate collective modes. Consequently, there is a single longitudinal Goldstone mode, which incorporates density fluctuations and can hybridize with particle--hole excitations, as we show below. To identify Goldstone modes, we define an effective translation vector as follows. We note that a uniform displacement $\vec u$ changes the Bragg peaks by $\delta\rho_{\vec G}=-i(\vec G\cdot\vec u)\rho_{\vec G}^{(0)}$, where $\vec G$ is a crystalline reciprocal lattice vector and $\rho_{\vec G}^{(0)}$ is the equilibrium amplitude. Collecting fluctuations associated with different Bragg peaks, we may write $\delta\vec \rho =B \vec u$ with $B_{\vec{G}_j}=-iG_{j}\rho^0_{\vec G}$. This motivates us to define the effective translation vector $\vec U_{\lambda\vq}=B^+\delta\vec\rho_{\lambda\vq}$, where $\delta\vec\rho_{\lambda\vq}$ represents the density fluctuations induced by mode $\lambda$ and $B^+$ denotes the pseudoinverse of $B$. We then characterize each mode by its translation weight and longitudinal polarization, 
\beq 
w_{\lambda\vq,\rm{tr}}=|\vec U_{\lambda\vq}|^2, \qquad f_{\lambda\vq,L}=|\hat{\vq}\cdot\vec U_{\lambda\vq}|^2/|\vec U_{\lambda\vq}|^2.
\eeq

\begin{figure}[pth!]
\centering
\includegraphics[width=0.99\linewidth]{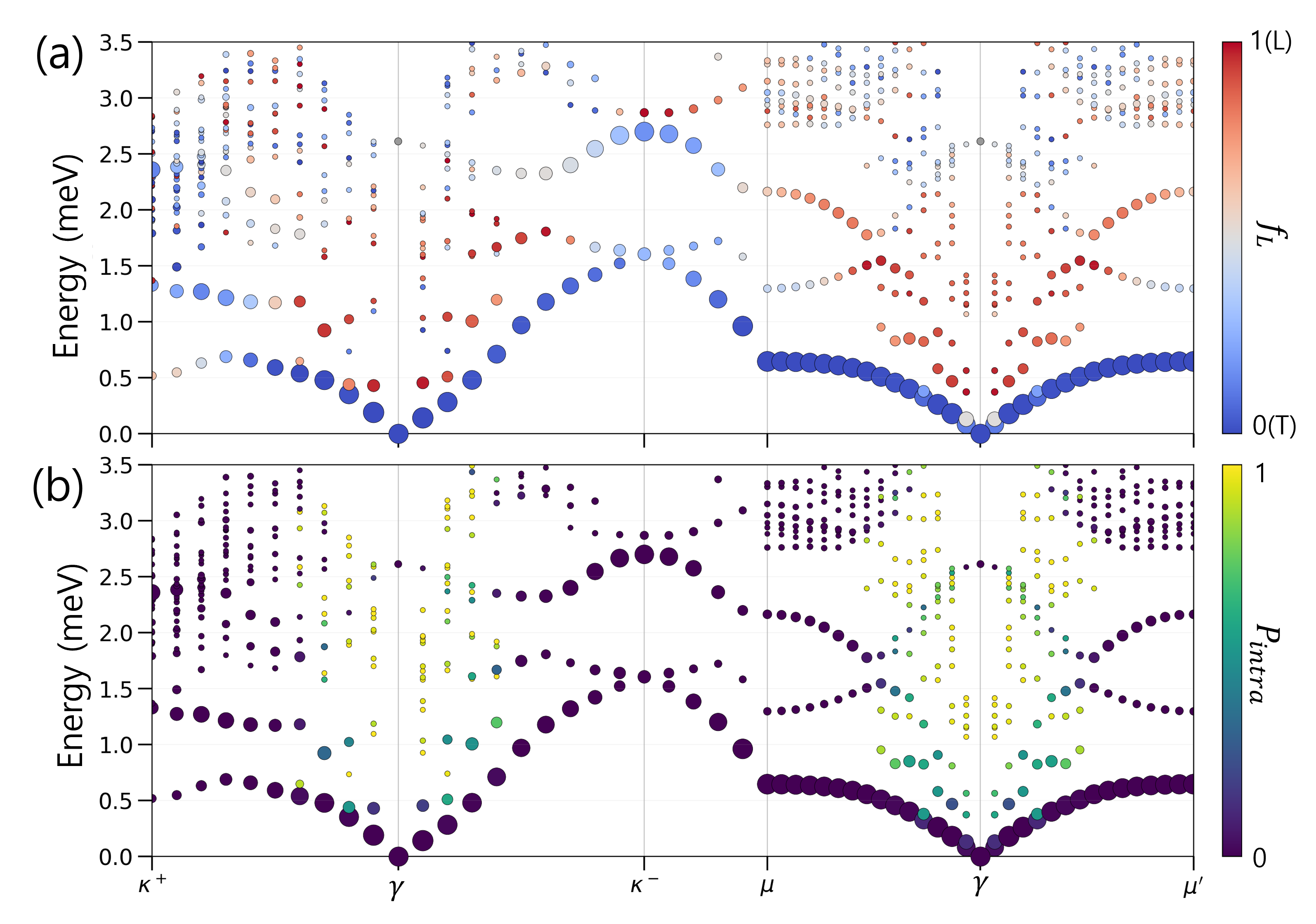}
\caption{\textbf{Collective excitations of the hole-doped mWC.}
(a) TDHF spectrum along a high-symmetry path in the crystalline BZ. Marker sizes indicate translation weights, and colors represent polarization. At $\gamma$, we identify two gapless Goldstone modes (only transverse mode is shown) and one gapped mode below the chosen cutoff $E_c=3.5\rm meV$. The gapped mode is shown in gray because polarization is ill-defined at $\vq=0$. (b) Same spectrum with colors representing the intraband particle--hole composition. Marker sizes are identical to those in (a).}
\label{fig:Collective}
\end{figure}

Fig.~\ref{fig:Collective}(a) shows the TDHF spectrum of the hole-doped mWC at $n=0.4\times10^{12}\mathrm{cm}^{-2}$, with marker sizes and colors indicating translation weights and polarization, respectively. While two gapless Goldstone modes (i.e. phonons) emerge from $\gamma$, only the transverse phonon retains a clear, strongly anisotropic dispersing branch. In contrast, the longitudinal translation weight is distributed among multiple modes. Fig.~\ref{fig:Collective}(b) shows the intraband particle--hole composition of each mode. The transverse phonon shows negligible intraband character, whereas several modes carrying longitudinal translation weight exhibit appreciable intraband contributions. These results suggest strong hybridization of the longitudinal phonon with the particle--hole continuum of the hole pocket. Consistent with this picture, the longitudinal mode becomes more clearly identifiable at a lower density where the hole pocket is smaller (see Supplemental Material \cite{si}). One possible explanation of this contrast is that longitudinal displacements induce smooth density fluctuations ($\vec G=0$) $\delta\rho_{\vq}\simeq-i\rho_0\vq\cdot \vec u$, whereas transverse displacements do not at leading order. However, the hybridization between the transverse phonon and particle--hole excitations may become stronger in the thermodynamic limit as the particle--hole continuum becomes denser.

\begin{figure}[pth!]
\centering
\includegraphics[width=0.99\linewidth]{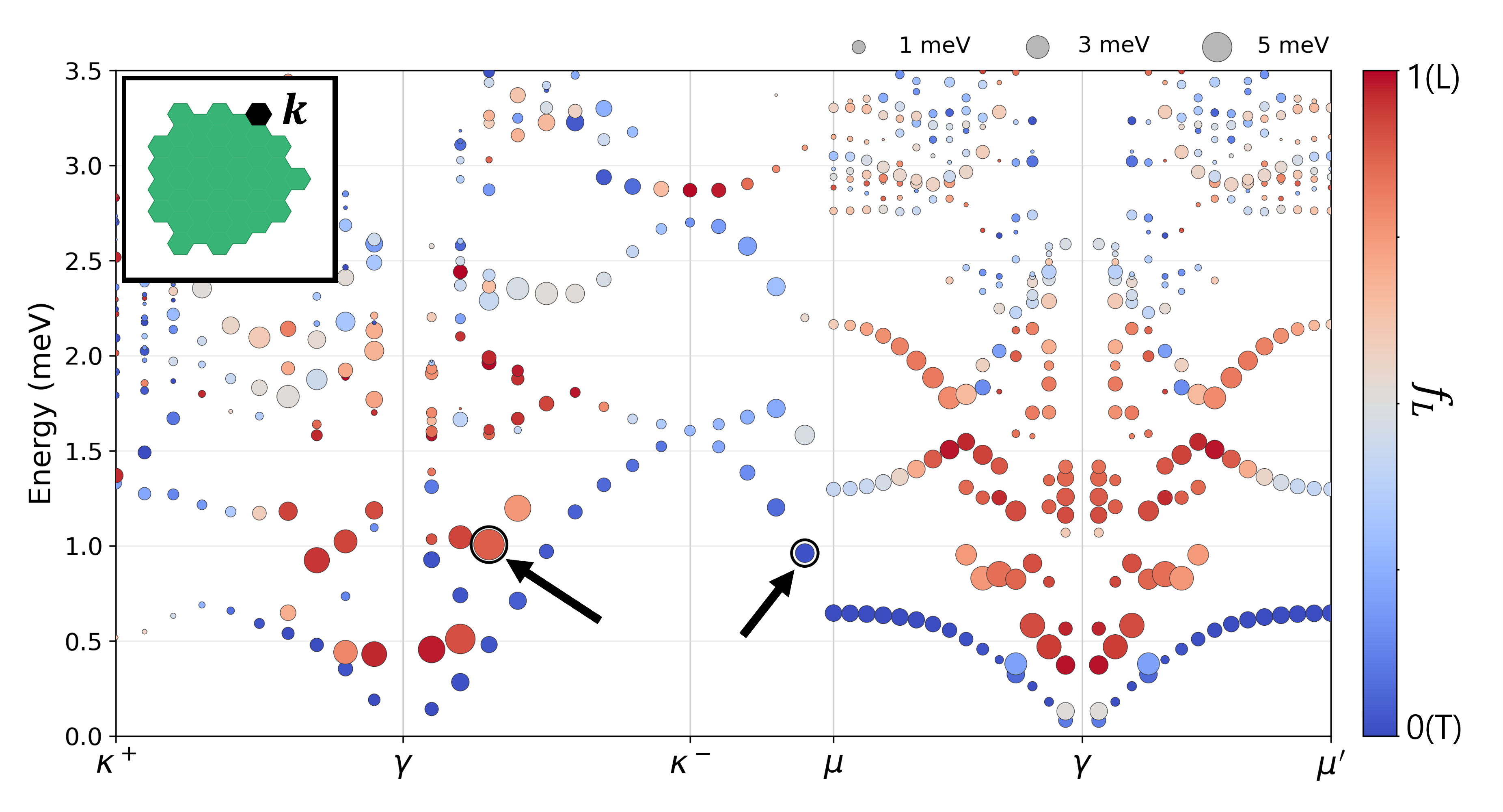}
\caption{\textbf{Quasiparticle coupling to TDHF excitations}. The same TDHF spectrum is shown where marker sizes represent the normalized quasiparticle coupling to each TDHF mode $\sqrt{N_c}|g^{\lambda}_{11}(\vk,\vq)|$ for a chosen representative $\vk$ at the Fermi edge (black point in the inset). Marker color again represents polarization. Black arrows show representative longitudinal-like and transverse-like modes with strong coupling strengths (5.42 meV and 2.06 meV, respectively).}
\label{fig:EPC}
\end{figure}

We now examine the coupling between the collective excitations and itinerant quasiparticles in the hole-doped mWC. The TDHF eigenvector $\vec x_{\lambda\vq}$ represents a density-matrix fluctuation $\delta P_{\lambda\vq}$. To linear order in the fluctuation, the HF
Hamiltonian is modified as
$\delta h_{\lambda\vq}
=h[P_0+\delta P_{\lambda\vq}]-h[P_0]$. Within the quasiboson approximation, quantizing each collective excitation gives
an effective coupling between quasiparticles and collective excitations,
\beq
H_{qp-ph}=
\sum_{\vk,\vq,m,n,\lambda}
\left[
g_{nm}^{\lambda}(\vk,\vq)
\gamma^\dg_{\vk+\vq,n}\gamma_{\vk m}
b_{\lambda \vq}
+\mathrm{H.c.}
\right],
\eeq
where the coupling vertex is
\beq
g_{nm}^{\lambda}(\vk,\vq)
=\sum_{\alpha\beta}
C_{\beta n}^*(\vk+\vq)
[\delta h_{\lambda\vq}]_{\vk,\beta\alpha}
C_{\alpha m}(\vk),
\eeq
and the TDHF eigenvectors are canonically normalized \cite{si}.

We focus on intraband coupling $g_{11}^{\lambda}(\vk,\vq)$ within the partially occupied HF band, fixing the initial quasiparticle momentum $\vk$ at the Fermi edge of the hole pocket. Fig.~\ref{fig:EPC} shows the TDHF spectrum with marker sizes proportional to the normalized coupling strength $\sqrt{N_c}|g_{11}^{\lambda}(\vk,\vq)|$, while colors again indicate polarization. The coupling exhibits strong mode dependence, and the magnitude is typically stronger for longitudinal-like modes ($\lesssim5.42$ meV) than transverse-like modes ($\lesssim2.06$ meV), consistent with the stronger hybridization between the longitudinal Goldstone mode and particle--hole excitations. These results demonstrate that, in the mWC, quasiparticle coupling cannot be characterized solely in terms of well-defined phonon branches, but should account for the full spectrum of collective excitations. While the mode dependence of the coupling is clear, its momentum dependence is affected by finite-mesh effects. In particular, the smallest accessible momentum is not sufficiently small to probe the long-wavelength limit, motivating future calculations at larger system sizes.

\textit{Outlook.}---
Our results demonstrate that itinerant carriers and crystalline order must be understood as coupled components of a single electronic state in mWCs. Energetically favorable self-doping does not guarantee stability against particle--hole fluctuations. Our finding is qualitatively consistent with transport experiments reporting mWCs only in a narrow hole-doped regime \cite{Han-Ju_26}. Determining the fate of unstable mWCs requires unrestricted microscopic calculations. Possible outcomes include crystals with reduced symmetry or qualitatively different charge patterns \cite{Abouelkomsan-Fu_arxiv26,Ge-Yu_arxiv26}.

Our analysis of collective excitations shows that the clearest phonon branch need not couple most strongly to quasiparticles. Understanding quasiparticle renormalization therefore would require investigation of the broader excitation spectrum. An interesting question is whether fluctuations near a particle--hole instability can lead to superconductivity observed near the crystalline state \cite{Han_Nature25}.

In conclusion, mWCs provide a setting in which Fermiology, crystalline order, and collective dynamics evolve together. Exploring a broader phase space in RMG varying layer number could help distinguish general features of mWCs from those specific to their microscopic electronic structure.

{\it Acknowledgement.---}
We thank Ziwei Wang and Chunli Huang for useful discussions. This work is supported by the Laboratory for Physical
Sciences through the Condensed Matter Theory Center.

\bibliography{refs}

\clearpage

\ifincludesupplement
  \clearpage
  \hypersetup{pageanchor=false}
  \def\CAPMAININCLUDESUPPLEMENT{1}
  \ifdefined\CAPMAININCLUDESUPPLEMENT
\def\finishsupplement{}
\else
\documentclass[aps,prl,twocolumn,floatfix,longbibliography,nofootinbib]{revtex4-2}
\usepackage[utf8]{inputenc}
\usepackage{natbib}
\usepackage{graphicx}
\usepackage{xcolor}
\usepackage[tbtags]{amsmath}
\usepackage{amssymb}
\usepackage{braket}
\usepackage{bm}
\usepackage{comment}
\usepackage{multirow}
\usepackage[normalem]{ulem}
\usepackage[colorlinks,allcolors=black]{hyperref}
\graphicspath{{../references/}{../figures/}{Figures/}{figures/}}
\renewcommand{\vec}[1]{\boldsymbol{#1}}
\def\vA{\vec{A}}
\def\vE{\vec{E}}
\def\vk{\vec{k}}
\def\vq{\vec{q}}
\def\vp{\vec{p}}
\def\vR{\vec{R}}
\def\vr{\vec{r}}
\def\ve{\varepsilon}
\def\dg{\dagger}
\def\nn{\nonumber}
\def\beq{\begin{eqnarray}}
\def\eeq{\end{eqnarray}}
\def\la{\langle}
\def\ra{\rangle}
\def\PP{{\cal P}}
\def\M{{\cal M}}
\def\Z{{\cal Z}}
\def\F{{\cal F}}
\def\E{{\cal E}}
\def\N{{\cal N}}
\begin{document}
\def\finishsupplement{\onecolumngrid\bibliography{refs}\end{document}}
\fi

\setcounter{page}{1}
\setcounter{section}{0}
\setcounter{equation}{0}
\setcounter{figure}{0}
\setcounter{table}{0}
\renewcommand{\thesection}{S\arabic{section}}
\renewcommand{\theequation}{\thesection.\arabic{equation}}
\renewcommand{\thefigure}{S\arabic{figure}}
\renewcommand{\thetable}{S\arabic{table}}
\renewcommand{\theHsection}{supp.\arabic{section}}
\renewcommand{\theHequation}{supp.\arabic{section}.\arabic{equation}}
\renewcommand{\theHfigure}{supp.\arabic{figure}}
\renewcommand{\theHtable}{supp.\arabic{table}}

\begin{widetext}
\begin{center}
{\bf Supplemental Material for \emph{Stability, Collective Modes, and Quasiparticle--Phonon Coupling of Metallic Wigner Crystals in Rhombohedral Graphene}}\\[3pt]
Sunghoon Kim, Yang-Zhi Chou, Jay D. Sau, and Sankar Das Sarma
\end{center}

In this Supplemental Material, we present the details of the HF calculations, the stability analysis, the TDHF calculations, and the coupling between quasiparticles and TDHF excitations. 

\refstepcounter{section}\setcounter{equation}{0}
\section{HF calculations in RMG}
\label{sec:supp_HF}

\subsection{Noninteracting RMG Hamiltonian}

We consider the tight-binding model of spin-valley-polarized rhombohedral tetralayer graphene \cite{Auerbach_NatPhys25}.
In the sublattice basis
$(A_1,B_1,A_2,B_2,A_3,B_3,A_4,B_4)$,
the noninteracting Hamiltonian is
\beq
H_0=\sum_{\vp}\Psi_{\vp}^\dg
h_{\rm RMG}(\vp)\Psi_{\vp},
\eeq
where $\Psi_{\vp}$ denotes the spinor combining layer and sublattice.
The Hamiltonian under a displacement field can be written as
\beq
h_{\rm RMG}(\vp)=
\begin{pmatrix}
h_1(\vp) & T_1(\vp) & T_2 & 0\\
T_1^\dg(\vp) & h_2(\vp) & T_1(\vp) & T_2\\
T_2^\dg & T_1^\dg(\vp) & h_3(\vp) & T_1(\vp)\\
0 & T_2^\dg & T_1^\dg(\vp) & h_4(\vp)
\end{pmatrix}.
\eeq
The intralayer and interlayer blocks are
\begin{align}
h_\ell(\vp)
&=
\begin{pmatrix}
0 & -t_0 f(\vp)\\
-t_0 f^*(\vp) & 0
\end{pmatrix}
+
\begin{pmatrix}
\varepsilon_{\ell A} & 0\\
0 & \varepsilon_{\ell B}
\end{pmatrix},
\\
T_1(\vp)
&=
\begin{pmatrix}
t_4 f(\vp) & t_3 f^*(\vp)\\
t_1 & t_4 f(\vp)
\end{pmatrix},
\qquad
T_2=
\begin{pmatrix}
0 & t_2/2\\
0 & 0
\end{pmatrix}.
\end{align}
Here $(t_0,t_1,t_2,t_3,t_4)=(3100,380,-15,290,141)\mathrm{meV}$ denote different dominant hoppings, and $f(\vp)=
\sum_{j=0}^{2}
\exp\left[
i\vp\cdot R_{2\pi j/3}
\begin{pmatrix}
0\\ a/\sqrt{3}
\end{pmatrix}
\right]$ is the structure factor of graphene, where $R_\theta$ is a rotation matrix in 2D
and $a=2.46\,\mathrm{\AA}$ is the graphene lattice constant. The perpendicular displacement field is modeled by a constant potential difference $U$ between adjacent layers, $u_\ell=\left(\ell-\frac52\right)U$ for $\ell=1,2,3,4$. We include the additional onsite potentials $\delta$ and
$\Delta_2$ to obtain layer-dependent onsite energies as
\begin{align}
(\varepsilon_{1A},\varepsilon_{1B})
&=(u_1+\Delta_2,u_1+\Delta_2+\delta),\nn\\
(\varepsilon_{2A},\varepsilon_{2B})
&=(u_2-\Delta_2+\delta,u_2-\Delta_2+\delta),\nn\\
(\varepsilon_{3A},\varepsilon_{3B})
&=(u_3-\Delta_2+\delta,u_3-\Delta_2+\delta),\nn\\
(\varepsilon_{4A},\varepsilon_{4B})
&=(u_4+\Delta_2+\delta,u_4+\Delta_2),
\end{align}
where $\delta=10.5\,\mathrm{meV}$ and
$\Delta_2=2\,\mathrm{meV}$.
Throughout this work, we choose $U=60\,\mathrm{meV}$.

\subsection{Interacting Hamiltonian and HF calculations}
Diagonalizing the noninteracting Hamiltonian gives band energies $\varepsilon_a^0(\vp)$ and spinors $|u_a(\vp)\rangle$. We project the density operator onto the retained conduction-band states,
\begin{equation}
\rho_{\vq}=\sum_{\vp,a,b}\langle u_a(\vp+\vq)|u_b(\vp)\rangle
c^\dagger_{\vp+\vq,a}c_{\vp b},
\end{equation}
with the screened Coulomb interaction $V(q)=e^2\tanh(qd)/(2\epsilon_0\epsilon_r  q)$.

For a triangular crystal, a parent state is labeled by a reduced momentum $\vk$ in the crystalline BZ and a composite index $\alpha=(a,\vec{G}_\alpha)$, with $\vec{G}_\alpha$ a reciprocal lattice vector. Its momentum in the original graphene BZ is $\vec{K}+\vk+\vec{G}_\alpha$. Then the density matrix is $P_{\alpha\beta}(\vk)=\langle d^\dagger_{\vk\alpha}d_{\vk\beta}\rangle$. Let us express the interaction in the basis labeled $I,J,K,L$ combining a reduced momentum and a folded-state label. Defining the antisymmetrized interaction $\bar V_{IK;JL}=V_{IK;JL}-V_{IK;LJ}$, the HF energy and field can be written as
\begin{align}
E_{\rm HF}[P]&=\sum_{IJ}(h_0)_{IJ}P_{IJ}
+\frac12\sum_{IJKL}\bar V_{IK;JL}P_{IJ}P_{KL},\label{eq:supp_Efunctional}\\
\delta E_{\rm HF}&=\sum_{IJ}h_{IJ}[P] \,\delta P_{IJ},\qquad
h[P]=h_0+\Sigma_{\rm H}[P]+\Sigma_{\rm F}[P].\label{eq:supp_energyvariation}
\end{align}

The energy is minimized self-consistently at fixed electron density $n$ and trial unit-cell density $n_{\rm uc}$, and then compared across trial $n_{\rm uc}$. The crystal filling is $\nu=n/n_{\rm uc}$, i.e. $\nu<1$ and $\nu>1$ correspond to net hole and electron self-doping of the commensurate WC, respectively. $n_{\rm uc}$ fixes the reciprocal lattice and hence the folded basis. We obtain the self-consistent solution at different $n_{\rm uc}$ while preserving $n$ and selecting the minimum HF energy. 

At a self-consistent solution $P_0$, the reconstructed HF eigenvectors $C_{\alpha m}(\vk)$ define
\begin{align}\label{eq:si_basis}
d_{\vk\alpha}&=\sum_m C_{\alpha m}(\vk)\gamma_{\vk m},\\
\sum_\beta h_{\alpha\beta}[P_0](\vk)C_{\beta m}(\vk)
&=\epsilon_m(\vk)C_{\alpha m}(\vk),\nonumber\\
(P_0)_{\alpha\beta}(\vk)&=\sum_m C^*_{\alpha m}(\vk)f_m(\vk)C_{\beta m}(\vk),\nonumber
\end{align}
where $f_m(\vk)$ is the occupation of the HF state. For numerical calculations, we use $U=60$ meV, $\epsilon_r=15$, $d=25$ nm and $N_c=N_k^2$ unit cells. The main results are obtained for $N_k=30$.

It is important to note that the optimized $n_{\rm uc}$ is the minimum within the triangular-lattice variational family, and cannot serve as a proof of stability against arbitrary-momentum particle--hole fluctuations. 

\refstepcounter{section}\setcounter{equation}{0}
\section{Stability analysis}\label{sec:supp_stab}

A saddle point density matrix $P_0$ is diagonal in the reconstructed HF basis, with energies $\epsilon_m(\vk)$ and occupations $f_m(\vk)$. Consider a momentum $\vq$ associated with particle--hole fluctuation, and the first-order density-matrix fluctuation $\delta P^{\vq}_{\vk,\alpha\beta}
=\delta\langle d^\dagger_{\vk\alpha}d_{\vk+\vq,\beta}\rangle$. Its components in the HF basis are
\begin{align}\label{eq:si_transform}
x_{\vk mn}(\vq)&=\delta\langle\gamma^\dagger_{\vk m}
\gamma_{\vk+\vq,n}\rangle,\\
\delta P^{\vq}_{\vk,\alpha\beta}&=\sum_{mn}
C^*_{\alpha m}(\vk)x_{\vk mn}(\vq)C_{\beta n}(\vk+\vq).\nonumber
\end{align}
We define the particle--hole excitation energy and occupation difference for the transition $\mu=(\vk,m,n)$:
\begin{equation}\label{eq:si_dD}
\Delta_\mu(\vq)=\epsilon_n(\vk+\vq)-\epsilon_m(\vk),\qquad
d_\mu(\vq)=f_m(\vk)-f_n(\vk+\vq).
\end{equation}
The same self-consistent HF functional determines the induced one-body field for any fluctuation,
\begin{equation}\label{eq:si_K}
\delta h^{\vq}=K_{\vq}\delta P^{\vq},\qquad
K_{\vq}=\left.\frac{\delta h}{\delta P}\right|_{P_0,\vq},
\end{equation}
where the kernel $K_{\vq}$ contains both direct and exchange variations of the interaction. Equivalently, we can use HF-band representation to write,
\begin{equation}
\label{eq:supp_Fexplicit}
(F_{\vq}x)_{\vk mn}
=
\sum_{\vk'rs}\bar V_{(\vk+\vq,n)(\vk',r);(\vk,m)(\vk'+\vq,s)}\,
 x_{\vk'rs}(\vq)
=\sum_{\alpha\beta}
C^*_{\beta n}(\vk+\vq)
[\delta h^{\vq}]_{\vk,\beta\alpha}C_{\alpha m}(\vk).
\end{equation}
Let us introduce $D_{\vq}$ and $\Delta_{\vq}$ to denote diagonal matrices for occupation and energy differences. Then we obtain the stability matrix
\begin{equation}\label{eq:si_S}
S(\vq)=D_{\vq}^{-1}\Delta_{\vq}+F_{\vq},
\end{equation}
acting on transitions with $d_\mu\neq0$. Let us denote the eigenvectors of the stability
matrix by $v_{\alpha\vq}$, satisfying
\beq
S(\vq)v_{\alpha\vq}
=s_\alpha(\vq)v_{\alpha\vq},
\qquad
v_{\alpha\vq}^\dagger v_{\beta\vq}
=\delta_{\alpha\beta}.
\eeq
Expanding the self-consistent HF functional yields
\beq
\delta E_{\rm HF}^{(2)}
=\frac{1}{2N_c}\sum_{\vq}
v_{\vq}^\dagger S(\vq)v_{\vq}
.
\eeq
A negative eigenvalue $s_\alpha(\vq)$ signifies an
energy-lowering particle--hole fluctuation.

Fig.~\ref{fig:si_stability} shows the lowest eigenvalues of the stability matrix of mWCs in the full crystalline BZ as a function of $n$ for both $N_k=24$ and $N_k=30$. Both cases reveal significant instability at high densities $(n\gtrsim0.42)$ when the mWC features electron pockets. In the vicinity of the high-density boundary of the hole-doped regime, the onset of instability can depend on system size.

\begin{figure*}[t]
\centering
\includegraphics[width=0.9\textwidth]{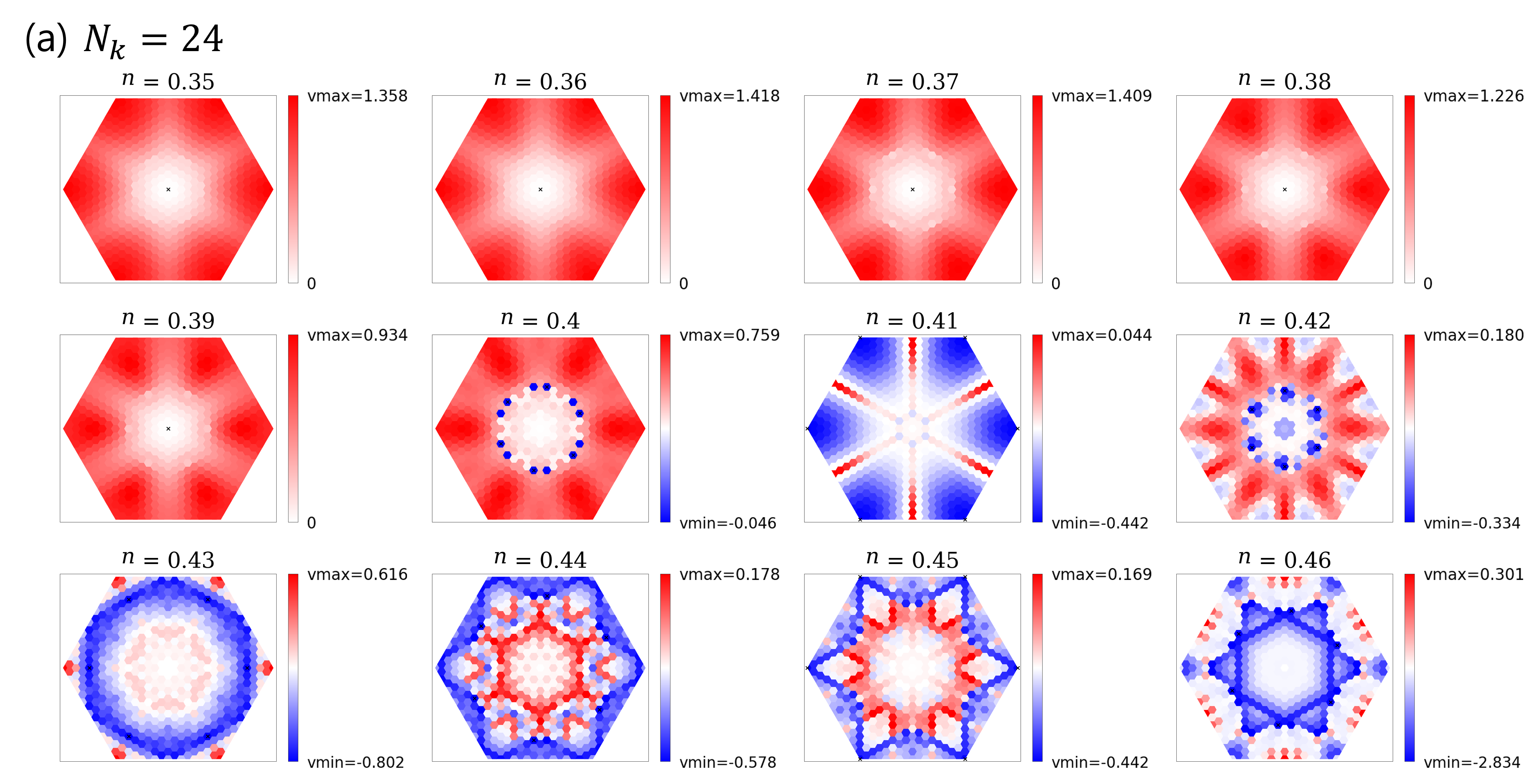}\\[3pt]
\includegraphics[width=0.9\textwidth]{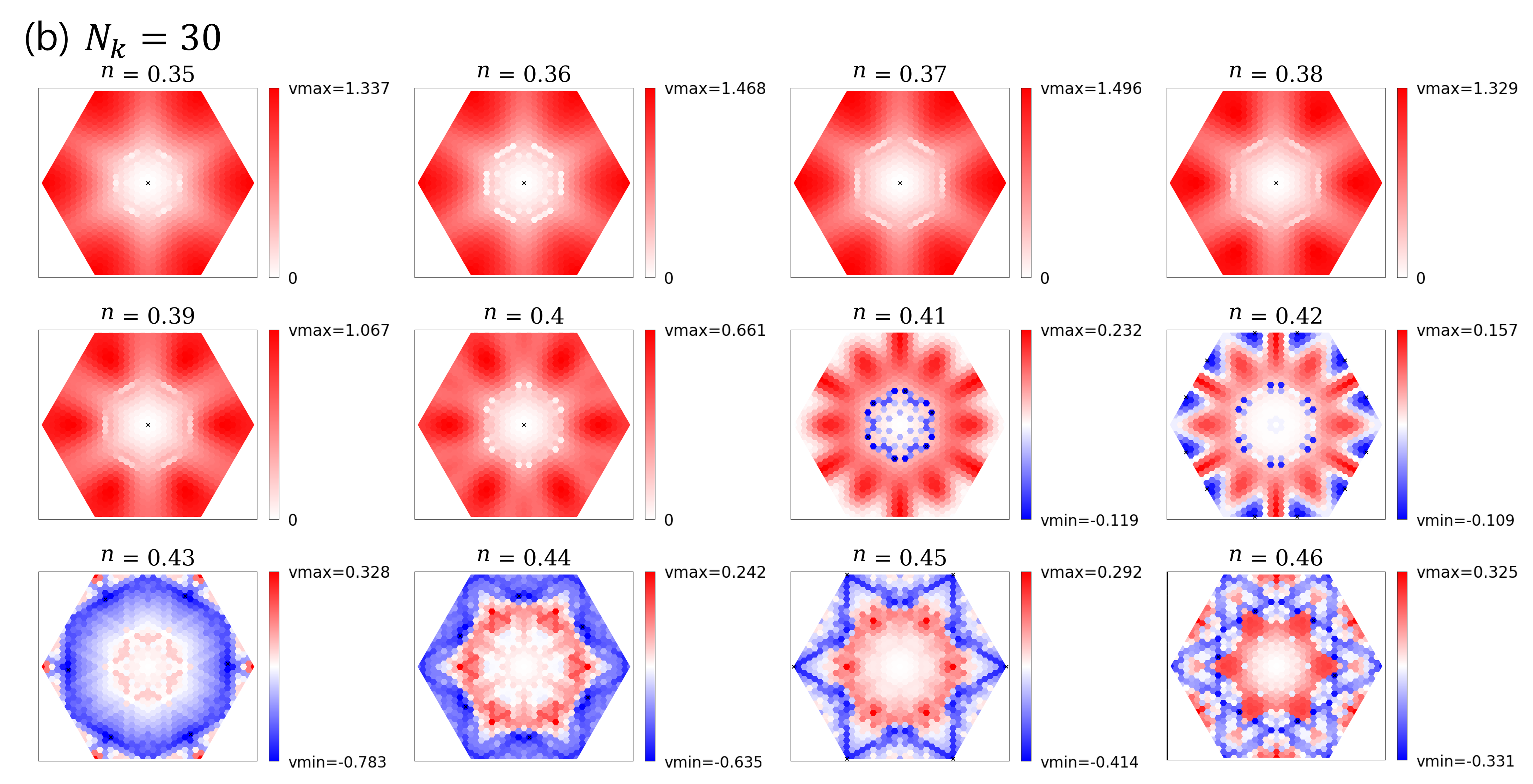}
\caption{\textbf{Density dependence of the lowest HF stability eigenvalue.} (a) and (b) show contour maps of $s_{\min}(\vq)$ of mWCs as a function of electron density for $N_k=24$ and $N_k=30$, respectively.}
\label{fig:si_stability}
\end{figure*}

\refstepcounter{section}\setcounter{equation}{0}
\section{TDHF calculations}\label{sec:supp_TDHF}

Linearizing the equations of motion of the density matrix gives the TDHF equation,
\begin{equation}\label{eq:si_L}
i\hbar\partial_t\vec{x}_{\vq}=L_{\vq}\vec{x}_{\vq},\qquad
L_{\vq}=\Delta_{\vq}+D_{\vq} F_{\vq},
\end{equation}
where the eigenmodes $\vec{x}_{\lambda\vq}$ satisfy
$L_{\vq}\vec{x}_{\lambda\vq}=E_\lambda(\vq)\vec{x}_{\lambda\vq}$. For a locally stable HF state, the eigenvalues are real.

The particle--hole bilinears are not canonical bosons. Writing $A_\mu(\vq)=\gamma_{\vk m}^\dagger\gamma_{\vk+\vq,n}$, the quasiboson approximation gives $\langle[A_\mu,A_\nu^\dagger]\rangle_0=d_\mu\delta_{\mu\nu}$. A mode creation operator can be constructed as
\begin{equation}\label{eq:supp_boson}
b_{\lambda\vq}^\dagger=\sum_\mu\frac{x_{\lambda\vq,\mu}}{d_\mu}\,A_\mu^\dagger(\vq),
\end{equation}
so that $\langle[b_{\lambda\vq},b_{\lambda'\vq}^\dagger]\rangle_0=\vec x_{\lambda\vq}^\dagger D_{\vq}^{-1}\vec x_{\lambda'\vq}$. Here, $\lambda$ labels positive-energy, positive-norm TDHF modes. We use the canonical TDHF normalization $\vec{x}_{\lambda\vq}^\dagger D_{\vq}^{-1}\vec{x}_{\lambda'\vq}
=\delta_{\lambda\lambda'}$.

\begin{figure*}[t]
\centering
\includegraphics[width=0.7\textwidth]{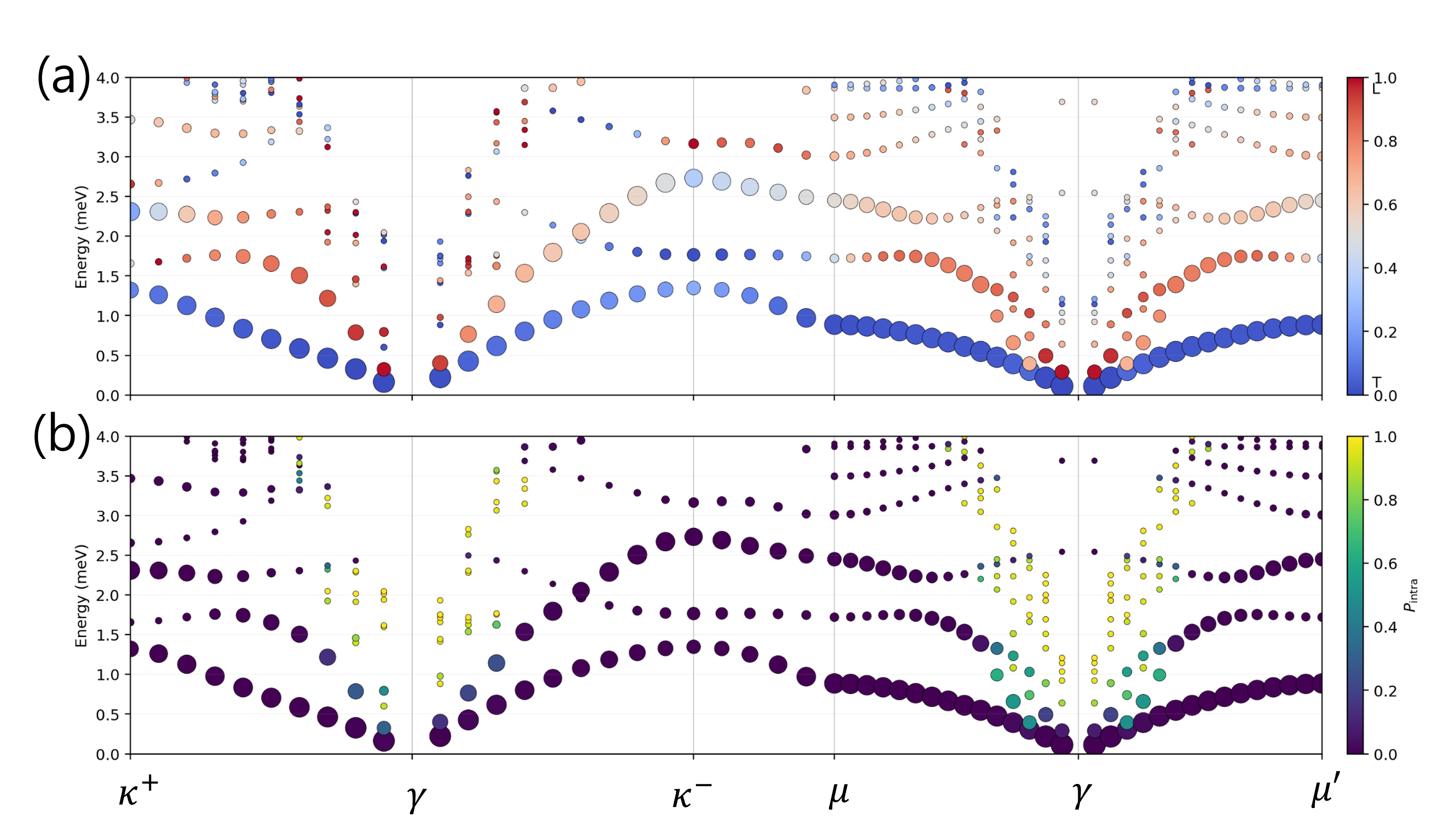}
\caption{\textbf{TDHF spectrum of the mWC with a smaller hole pocket ($n=0.34\times10^{12}\mathrm{cm}^{-2}$).}
Marker sizes indicate translation weights, while marker colors in (a) and (b) represent polarization and intraband particle--hole composition, respectively.}
\label{fig:si_tdhf_n0.34}
\end{figure*}

Fig.~\ref{fig:si_tdhf_n0.34} shows the TDHF spectrum at a lower density
$n=0.34\times10^{12}\mathrm{cm}^{-2}$, where the hole pocket is smaller
than the mWC at $n=0.4\times10^{12}\mathrm{cm}^{-2}$ in the main text. The transverse phonon remains well defined at both densities.
At the lower density, the longitudinal phonon becomes more identifiable as a dispersing branch along $\gamma\rightarrow\kappa^{-}$ and $\gamma\rightarrow\kappa^{+}$, but again strongly mixes with particle--hole excitations along other directions. This behavior indicates reduced, momentum-dependent hybridization with intraband particle--hole excitations as the hole pocket gets smaller.

A canonically normalized TDHF eigenmode $\vec{x}_{{\lambda}\vq}$ determines the fluctuation $\delta P^{\vq}_{\lambda}$ through Eq.~\eqref{eq:si_transform}. The induced perturbation to linear order is
\begin{equation}\label{eq:si_dh}
\delta h^{\vq}_{\lambda}=K_{\vq}\delta P^{\vq}_{\lambda}
=h[P_0+\delta P^{\vq}_{\lambda}]-h[P_0]
\quad\text{(linear order)}.
\end{equation}
The matrix element between reconstructed HF states defines the mode-resolved coupling vertex,
\begin{equation}\label{eq:si_g}
g^{\mathrm \lambda}_{nm}(\vk,\vq)=\sum_{\alpha\beta}
C^*_{\beta n}(\vk+\vq)
[\delta h^{\vq}_{\lambda}]_{\vk,\beta\alpha}
C_{\alpha m}(\vk).
\end{equation}

\end{widetext}

\finishsupplement

\fi

\end{document}